\documentstyle[pra,aps,twocolumn,epsfig]{revtex}

\begin{document}

\title{Collisional effects on the collective laser cooling of trapped bosonic gases}

\author{Luis Santos and Maciej Lewenstein}

\address{Institut f\"ur Theoretische Physik, Universit\"at Hannover,
 Appelstr. 2, D--30167 Hannover,
Germany}

\maketitle

\begin{abstract}
We analyse the effects of atom--atom collisions on collective laser cooling scheme. 
We derive a quantum Master equation which describes the laser cooling in presence of 
atom--atom collisions in the weak--condensation regime. Using such equation, we perform 
Monte Carlo simulations of the population dynamics in one and three dimensions. We observe 
that the ground--state laser--induced condensation is maintained in the presence of collisions. 
Laser cooling causes a transition from a Bose--Einstein distribution 
describing collisionally induced equilibrium,to a distribution 
with an effective zero temperature. We analyse also the effects of atom--atom collisions 
on the cooling into an excited state of the trap.

\end{abstract}
%\narrowtext

\section{Introduction}
\label{sec:Intro}

In the recent years, laser cooling has constituted one of the most active research fields in 
atomic physics \cite{Nobel}. However, the laser cooling techniques by themselves have not 
allowed to reach temperatures for which the quantum statistical effects become evident. 
In particular, only the combination of laser cooling, and evaporative \cite{BEC} or 
sympathetic cooling \cite{Symp} has permited in the last years to observe experimentally, 
the Bose--Einstein condensation (BEC) in alkali gases, seventy years after its theoretical 
prediction \cite{Bose24}. The question whether it is or it is not possible to achieve the 
BEC only with laser cooling techniques remain, at least as an intellectual challenge. The 
laser--induced BEC is, however, not only an academic problem, but has several advantages with 
respect to the 
nowadays widely--employed collisional mechanisms (as evaporative cooling). These 
advantages are: (i) the number 
of atoms does not decrease during the cooling process; (ii) it is possible to design a 
non--destructive BEC detection by fluorescence measurements; (iii) reacher effects 
can appear, since now the system is open, i.e. it is not in thermal equilibrium; 
(iv) laser--induced condensation can be used to design 
techniques to pump atoms into the condensate. The latter is specially important in the contex of 
future atom--laser devices \cite{Spreeuw95,Janicke96,Janicke99,Bar}.

The main problem which prevents experimentalists from obtaining BEC by optical means 
is the reabsorption of 
spontaneously emitted photons. The most effective laser--cooling techniques (such as VSCPT 
\cite{VSCPT}, or Raman  cooling \cite{Raman}), are based on the crucial concept of dark states 
\cite{Orriols}, i.e. states which cannot absorb the laser light, but can receive 
population via incoherent pumping, i.e. via spontaneous emission. Unfortunately, the atoms 
occupying the dark states are not unaffected by the photons spontaneously emitted by other 
atoms. This problem turns to be very important at high densities, as those required for the 
BEC \cite{Reabproblem}; in such conditions dark--state cooling techniques cease 
to work adequately. Several 
remedies to the reabsorption problem have been proposed, as the reduction of the dimensionality of 
the trap from three to two or one dimensions \cite{Reabproblem}, or the use of traps 
with frequencies, $\omega$, of the order
 of the recoil frequency ($\omega_R=\hbar k_L^2/2M$, where $k_L$ is the laser wavevector and M is 
the atomic mass) \cite{Janicke96}. Other, perhaps more promising, idea consists in exploiting the 
dependence of the 
reabsorption probability on the fluorescence rate $\gamma$. In particular, in the so--called 
{\em Festina Lente} limit \cite{Festina}, when $\gamma <\omega$ with $\omega$ the trap frequency,
 the heating effects of the reabsorption can be neglected. Another proposal consists in 
working in the so--called Bosonic Accumulation Regime \cite{Bar}, in which the reabsorption 
can, under certain conditions, even help to build up the condensate. In the following we shall assume 
that the considered system fulfills the Festina Lente limit.

In a series of papers \cite{1Atom,Manyatoms}, we have proposed a cooling mechanism (which we have called 
Dynamical cooling) which permits the cooling of an atomic sample into an arbitrary single state of an 
harmonic trap, beyond the Lamb--Dicke limit (i.e. when the Lamb--Dicke parameter $\eta>1$, with 
$\eta^2=\omega_R/\omega$). 
The cooling mechanism employs 
laser pulses of different frequencies (and eventually different directions, phases and intensities),
in such a way that a particular state of the trap remains dark during the cooling process, acting 
as a trapping state. Therefore, the population is finally transferred to this particular state. We 
have first analysed the particular situation of a single atom in the trap \cite{1Atom}, and 
extended the analysis to a collection of trapped bosons \cite{Manyatoms}. We have shown that the 
bosonic statistics helps to achieve more robust and rapid condensation, as well as to produce 
non--linear effects, such as hysteresis and multistability phenomena.

	However, all the calculations performed so far in the analysis of the dynamical cooling 
scheme do not take into account the atom--atom collisions, i.e. are considered in the so--called 
{\em ideal gas} limit. The ideal gas limit imposes important restrictions to the physical system, in 
particular the atomic density cannot be very large. An interesting possibility in order to 
achieve quasi--ideal gases consists in the ``switching--off'' of the $s$--wave scattering length 
$a$ (which is the main contribution to the atom--atom collisions for sufficiently low energies), 
either by employing magnetic fields (tuning the so--called Feshbach resonances \cite{Ketterle}) 
or by using a red--detuned laser tuned between molecular resonances as proposed by 
Fedichev {\it et al} \cite{Fedichev}. However, without special precautions, the effects 
of the atom--atom collisions play a 
substantial role. It is the aim of this paper to analyse such effects in the context of our 
dynamical laser cooling scheme.

	In recent years, C. W. Gardiner, P. Zoller and collaborators have devoted a series of 
papers \cite{QK1,QK2,QK3,QK4,QK5} to the decription of interacting Bose gases with and without 
trapping potentials. These authors have developed a quantum kinetic theory 
of Bose gases. In particular, for the case of a weakly 
interacting gas, a so--called Quantum Kinetic Master Equation (QKME) has been formulated 
\cite{QK1}, which is a quantum stochastic equation for the kinetics of the dilute Bose gas, that 
describes the behavior and formation of the condensate. This equation is very difficult to 
simulate, and therefore various simplifications have been proposed. 
Particularly interesting results are obtaining by using the 
so--called Quantum Boltzmann Master Equation (QBME) which neglects all spatial 
inhomogeinity of the trapped states \cite{QK2}. 
Although this is an extreme simplification, the (very much easier) simulation of the QBME give a 
good idea of the solutions that the QKME could produce. In the following, we shall show that the 
master equation (ME) which describes the laser cooling problem in the presence of
 atom--atom collisions can be, 
in the case of the weak--condensation regime, splitted into two independent parts, one accounting 
for the collisional effects (which has the form of the QBME proposed in Ref.\ \cite{QK1}), and 
another which describes the laser cooling process, and has the form of the ME already developed for 
the case without collisions \cite{Manyatoms}.

	The structure of the paper is as follows. In Sec.\ \ref{sec:Model}, we derive the 
quantum ME which describes the laser cooling plus collisions in the weak--condensation regime. 
In Sec.\ \ref{sec:1D}, we present the results for one--dimensional excited--state 
cooling. Sec.\ \ref{sec:3D} is devoted to the three--dimensional
results for the case of ground--state cooling. Here, we use additional ergodic 
approximation which assures fast redistribution of atoms within an energy shell. 
Finally, in Sec.\ \ref{sec:conclu} we summarize some conclusions.

\section{Model. Master Equation.}
\label{sec:Model}

We assume in this paper the same atomic model as that presented in Refs.\ \cite{1Atom,Manyatoms}, 
i.e. a three--level $\Lambda$--system, composed of a ground--state level $|g\rangle$, a
metastable state $|e\rangle$ and an auxiliary third fast--decaying state $|r\rangle$. Two lasers
excite coherently the resonant Raman transition
$|g\rangle\rightarrow|e\rangle$ (with an associated effective Rabi frequency
$\Omega$), while the repumping laser in or off--resonance with the transition
$|g\rangle\rightarrow|r\rangle$ pumps optically the atom into $|g\rangle$. With this three level
scheme, one obtains an effective two--level system with an effective spontaneous emission rate
$\gamma$, which can be easily controlled by varying the intensity or the detuning of the repumping 
laser \cite{Marzoli94}. 
In the following we follow the same notation as in Refs.\ 
\cite{Manyatoms}. Let us introduce the annihilation and creation operators of atoms in the ground
(excited) state and in the trap level $m$ $(l)$, which we will call $g_{m}$, $g_{m}^{\dag}$ ($e_{l}$,
$e_{l}^{\dag}$). These operators fulfill the bosonic commutation relations 
$[g_{m},g_{m'}^{\dag}]=\delta_{mm'}$ and $[e_{l},e_{l'}^{\dag}]=\delta_{ll'}$. 
Using standard methods of the theory of quantum stochastic processes 
\cite{Gardinerbook1,Gardinerbook2,Carmichaelbook,Carmichaelbooknew} one can develop the quantum ME which 
describes the atom dynamics \cite{Manyatoms}
\begin{equation}
\dot\rho(t)={\cal L}_0\rho+{\cal L}_1\rho+{\cal L}_2\rho,
\label{ME}
\end{equation}
where
\begin{mathletters}
\begin{eqnarray}
{\cal L}_0\rho&=&-i\hat H_{eff}\rho(t)+i\rho(t)\hat H_{eff}^{\dag}+{\cal J}\rho(t), \\
{\cal L}_1\rho&=&-i[\hat H_{las},\rho(t)], \\
{\cal L}_2\rho&=&-i[\hat H_{col},\rho(t)], 
\end{eqnarray}
\end{mathletters}
with
\begin{mathletters}
\begin{eqnarray}
\hat H_{eff}&=&\sum_{m}\hbar\omega_{m}^{g}g_{m}^{\dag}g_{m}+\sum_{l}\hbar(\omega_{l}^{e}
-\delta)e_{l}^{\dag}e_{l} \nonumber \\
&-&i\hbar\gamma\int d\phi d\theta sin\theta {\cal W}(\theta,\phi) \nonumber \\
&\times& \sum_{l,m}|\eta_{lm}(\vec k)|^{2}e_{l}^{\dag}g_{m}g_{m}^{\dag}e_{l}, \\
\hat H_{las}&=&\frac{\hbar\Omega}{2}\sum_{l,m}\eta_{lm}(k_{L})e_{l}^{\dag}g_{m}+H.c.,\\
\hat H_{coll}&=&\sum_{m_1,m_2,m_3,m_4}\frac{1}{2}U_{m_1,m_2,m_3,m_4}
g_{m_4}^{\dag}g_{m_3}^{\dag}g_{m_2}g_{m_1},\\
{\cal J}\rho(t)&=&2\hbar\gamma\int d\phi d\theta \sin\theta{\cal W}(\theta,\phi) \nonumber \\
&\times& \sum_{l,m}[\eta_{lm}^{\ast}(\vec k)g_{m}^{\dag}e_{l}]\rho(t)[\eta_{lm}(\vec
k)e_{l}^{\dag}g_{m}].
\end{eqnarray}
\end{mathletters}
where $2\gamma$ is the single--atom spontaneous emission rate, $\Omega$ is the Rabi frequency 
associated with the atom transition and the laser field, 
$\eta_{lm}(k_{L})=\langle e,l|e^{i\vec k_{L}\cdot\vec r}|g,m\rangle$ are the Franck--Condon
factors,  ${\cal W}(\theta,\phi)$ is the fluorescence dipole pattern, $\omega_m^g$ 
($\omega_l^e$) are the energies of the ground (excited) harmonic trap level $m$ ($l$), and $\delta$ is 
the laser detuning from the atomic transition. The new term respect to what is considered in 
Refs.\ \cite{Manyatoms}, is that of $H_{coll}$, which describes the two--body interactions in the 
Bose gas. Only ground--ground collisions are considered because we assume that the laser 
interaction is sufficiently weak to guarantee that only few atoms are excited (formally 
we consider only one). In the regime we want to study, only $s$--wave scattering is important, 
and then:
\begin{equation}
U_{m_1,m_2,m_3,m_4}=\frac{4\pi\hbar^2a}{m}\int_{R^3}d^3x
\psi_{m_4}^{\ast}\psi_{m_3}^{\ast}\psi_{m_2}\psi_{m_1},
\end{equation}
where $\psi_{m_j}$ denotes the harmonic oscillator wavefunctions and $a$ denotes the 
scattering length. Following the simplifications of 
the QBME \cite{QK1,QK2}, we exclude the spatial dependence and therefore no transport or 
wave--packet spreading terms appear in (\ref{ME}).

In the following we are going to work in the so--called weak--condensation regime, where  no 
mean--field efects are considered. This means that we consider that the typical energy provided by 
the collisions is smaller than the oscillator energy. As shown in \cite{QK2}, in typical 
experiments this condition requires that the condensate cannot contain more than $1000$ particles. 
We shall work thus below such limit. We shall also consider that $\Omega\ll\omega$. Also, due to 
the Festina--Lente requirements, $\Omega$ is in general smaller than the typical collisional 
energy. Therefore we can formally establish the hierarchy 
${\cal L}_0\gg{\cal L}_2>{\cal L}_1$.

Let us define a projector operator ${\cal P}$:
\begin{equation}
{\cal P}X=\sum_{\vec n}|\vec n\rangle\langle\vec n|\langle\vec n|X|\vec n\rangle,
\label{Proy}
\end{equation}
and its complement ${\cal Q}=1-{\cal P}$, and 
$|\vec n\rangle\equiv |N_{0},N_{1},\dots;g\rangle$$\otimes
|0,0,\dots;e\rangle$ are the ground state configurations with $N_{j}$ atoms 
in the $j$--th level, and no excited atoms. It is easy to prove that:
\begin{equation}
{\cal L}_0{\cal P}={\cal PL}_1{\cal P}={\cal PL}_2{\cal P}=0.
\label{prop}
\end{equation}
Projecting the ME (\ref{ME}), one obtains:
\begin{mathletters}
\begin{eqnarray}
&& \dot v={\cal P}({\cal L}_{0}+{\cal L}_{1}+{\cal L}_{2})w, \label {vdot} \\
&& \dot w={\cal Q}({\cal L}_{0}+{\cal L}_{1}+{\cal L}_{2})v+
{\cal Q}({\cal L}_{1}+{\cal L}_{2})w, \label{wdot}
\end{eqnarray}
\end{mathletters}
where $v={\cal P}\rho$ and $w={\cal Q}\rho$. Laplace transforming ($v(t)\rightarrow\tilde v(s)$)
and solving the system of equations (assuming 
for simplicity $w(0)=0$), one obtains:
\begin{eqnarray}
s\tilde v(s)-v(0)&=&{\cal P}({\cal L}_{0}+{\cal L}_{1}+{\cal L}_2) \nonumber \\
&\times&[s-{\cal Q}({\cal L}_{0}+{\cal L}_{1}+{\cal L}_{2})]^{-1}
({\cal L}_{1}+{\cal L}_2)\tilde v(s).
\label{lapl}
\end{eqnarray} 
Performing the inverse Laplace transform, using (\ref{prop}) and 
${\cal P}\exp[-{\cal L}_0\tau]{\cal L}_1{\cal P}=
{\cal P}\exp[-{\cal L}_0\tau]{\cal L}_2{\cal P}=0$, and applying the Markov approximation 
following Ref.\ \cite{QK1}, the ME becomes up to order ${\cal O}({\cal L}_1^2)$:
\begin{equation}
\dot v(t)={\cal L}_{coll}v(t) + {\cal L}_{cool}v(t)
\label{ME2}
\end{equation}
where
\begin{equation}
{\cal L}_{coll}=-{\cal PL}_2{\cal L}_0{\cal L}_2
\label{Lcoll}
\end{equation}
describes the collisional part. In principle in ${\cal L}_{coll}$ appears a second term coming 
from the term between brackets in Eq.\ (\ref{lapl}), but since we consider the weak--condensation 
regime, we can employ the Born approximation as in Ref.\ \cite{QK1} to neglect these terms in the 
collisional part. Therefore the collisional part is described by a QBME as that of Refs.\ 
\cite{QK1,QK2}. The laser--cooling dynamics is described by
\begin{eqnarray}
{\cal L}_{cool}&=&-{\cal PL}_{1}[{\cal L}_{0}]^{-1}{\cal L}_{1} \nonumber \\
&+&\frac{1}{2}{\cal PL}_0\int_{0}^{\infty}d\tau\int_{0}^{\infty}d\tau' e^{-{\cal
L}_{0}(\tau-\tau')}{\cal L}_{1}e^{-{\cal L}_{0}\tau'}{\cal L}_{1},
\label{Lcool}
\end{eqnarray}
which has the same form of the ME calculated for the case of the laser cooling without 
collisions \cite{Manyatoms}.

Summaryzing, the dynamics of the system splits into two parts, (i) collisional part, 
described by a QBME, and (ii) laser--cooling part, described by the same ME as without collisions. The first 
correction to such splitting between both dynamics is of the order ${\cal L}_2{\cal L}_1^2$; 
therefore the independence between the collisions and laser cooling dynamics is only valid in 
the weak--interaction regime.

The independence of both dynamics, allows for an easy simulation of the laser cooling 
in presence of collisions. In particular, we simulate both dynamics using Monte Carlo methods, 
combining the numerical method of Ref.\ \cite{QK2}, with the simulations already presented in 
Refs.\ \cite{Manyatoms}. In the following two sections we shall present the results for 
one--dimensional and three--dimensional simulations respectively.

\section{One--dimensional results}
\label{sec:1D}

This section analyzes the case of a one dimensional harmonic trap, and it is mainly devoted 
to the  analysis of the laser cooling into states different that the ground 
state of the trap
 \cite{1Atom,Manyatoms} (the ground--state case in analysed in the next section for the more 
interesting case of three dimensions). The laser--cooling into an  
excited state of the trap is only calculated in the one--dimensional case, because 
its analysis becomes very complicated in higher dimensions. 
The reason for that is that the ergodic approximation, which we employ in Sec.\ \ref{sec:3D}, 
is incompatible with the analysis of the cooling in excited states of the trap. 

In this section 
we shall show that, as expected, the excited--state cooling is strongly affected by the collisions, 
even for modified low scattering lengths, in particular because the collisions act as a mechanism to 
empty the desired excited state, and therefore compete with the cooling mechanism which 
tends to populate such state. For usual scattering legths and atom densities,
the collisional processes are 
much faster than the typical cooling time, and therefore the excited state cooling is completely 
suppressed. However, the use of Feshbach resonances 
\cite{Ketterle}, or lasers \cite{Fedichev}, can modify the scattering length, in 
such a way that the typical time between collisions can be comparable with the typical cooling time. 
In this section,  we study the population dynamics for different modified scattering lengths, 
analysing the transition from an ideal--gas regime (with $a=0$ as that studied 
in Refs.\ \cite{Manyatoms}), to a regime in which atom--atom collisions constitute the 
dominant process.

\subsection{Collisional probabilities}

Following Ref.\ \cite{QK2} (Eq. (14)), the probability of a collision between two atoms 
respectivelly in the states $n_1$ and $n_2$ of an harmonic trap (of frequency $\omega$), 
to produce two atoms in the states 
$n_3$ and $n_4$ respectively, is given by:
\begin{eqnarray}
&&P_{coll}(n_1,n_2\rightarrow n_3,n_4)= \nonumber \\
&&\frac{4\pi}{\hbar^2\omega}|U_{n_1,n_2,n_3,n_4}|^2 \nonumber \\
&&\times N_{n_1}(N_{n_2}-\delta_{n_1,n_2})(N_{n_3}+1)(N_{n_4}+1+\delta_{n_3,n_4}),
\label{P1234}
\end{eqnarray} 
where $N_n$ denotes the occupation number of the level $n$ of the trap. 
The Kronecker deltas in the previous expression account for the bosonic factors appearing 
when two atoms in a particular level are created, or destructed. Considering a highly 
anisotropic trap of frequencies $\omega_x=\omega_y=\lambda\omega$, 
$\omega_z=\omega$, the problem becomes one--dimensional, and the expression (\ref{P1234}) 
takes the form
\begin{eqnarray}
&&P_{coll}(n_1,n_2\rightarrow n_3,n_4)= \nonumber \\
&&\xi
\omega 
\Delta_c(n_1n,n_2,n_3)\delta(n_1+n_2-n_3-n_4) \nonumber \\
&&\times N_{n_1}(N_{n_2}-\delta_{n_1,n_2})(N_{n_3}+1)(N_{n_4}+1+\delta_{n_3,n_4}),
\label{pcoll}
\end{eqnarray}
where the Dirac $\delta$ accounts for the energy conservation. In the previous expression:
\begin{eqnarray}
&&\Delta_c(n_1,n_2,n_3)=\left [2^{2(n_1+n_2)}n_1!n_2!n_3!(n_1+n_2-n_3)!\right ]^{-1} \nonumber \\
&& \left [\int_{-\infty}^{\infty}dzH_{n_1}(z)H_{n_2}(z)H_{n_3}(z)H_{n_1+n_2-n_3}(z)
e^{-2z^2} \right ] ^2,
\end{eqnarray}
where $H_j(z)$ is the Hermite polynomial of order $j$, and 
\begin{equation}
\xi=\frac{16\lambda^2}{\pi} \left ( \frac{a}{a_{HO}} \right )^2, 
\end{equation}
with $a_{HO}=\sqrt{\hbar/m\omega}$. For typical experimental situations $a/a_{HO}\sim 10^{-3}$, so 
$\xi=5\times 10^-6 r^2$, where $r=\lambda a/a_0$, where $a_0$ is the scattering 
length without any external modification. Therefore $r=0$ accounts for an ideal gas, whereas 
$r=\lambda$ accounts for an unmodified scattering length.

\subsection{Laser--cooling transition probabilities}

The probability of laser--induced transition of an atom from the state $n_1$ to the state $n_2$ 
of the trap is calculated in Refs.\ \cite{Manyatoms}, and it is given by
\begin{equation}   
P_{cool}(n_1\rightarrow n_2)=\frac{\Omega^2}{2\gamma}\Delta_l(n_1,n_2)N_{n_1}(N_{n_2}+1),
\label{pcool}
\end{equation}
where
\begin{eqnarray}
\Delta_l(n_1,n_2)&=&\frac{\Omega^{2}}{2\gamma}\int_{0}^{2\pi}d\phi\int_{0}^{\pi}
d\theta\sin\theta{\cal W}(\theta,\phi) \nonumber \\ 
&\times& 
\left
|\sum_{l}\frac{\gamma\eta_{ln_2}^{\ast}(\vec
k)\eta_{ln_1}(k_{L})}{[\delta-\omega(l-n_1)]+i\gamma R_{n_1l}}\right |^{2}.
\label{Gnm} 
\end{eqnarray}
with 
\begin{eqnarray}
R_{n_1l}&=&\int_{0}^{2\pi}d\phi\int_{0}^{\pi}d\theta\sin\theta{\cal W}(\theta,\phi) \nonumber \\
&\times& \sum_{n'}|\eta_{ln'}(\vec k)|^{2}(N_{n'}+1-\delta_{n',n_1})
\label{Rml}
\end{eqnarray}

\subsection{Dynamical cooling}

Let us briefly review at this point the dynamical cooling scheme 
proposed in refs.\ \cite{1Atom,Manyatoms}. It is easy to observe from
the form of the rates (\ref{Gnm}), and from the same arguments as those used in Ref.\ \cite{1Atom},
that as in the single--atom case two different dark--state mechanisms can be employed:
\begin{itemize}
\item "Franck--Condon"--dark--states. Let us assume a laser pulse with detuning $\delta=s\omega$
respect to the atomic transition, where $s$ is an integer number. It can be easily proved
\cite{1Atom} that a particular level of the trap $|m\rangle$ remains unemptied (dark) if the
Franck--Condon factor $\langle m+s|\exp(ikx)|m\rangle$ vanishes. In particular, the dark--state 
condition for $n=1$ is
\begin{equation}
\eta^{2}=s+1; \label{dcond1}
\end{equation}
\item "Interference"--dark--states. These dark--state mechanism is characteristic for dimensions
higher than one. Let us assume the two--dimensional problem, in which we have two orthogonal
lasers characterised by two different Rabi frequencies: $\Omega$ in direction $x$, and $A\Omega$ in
direction $y$. The factor $A$ indicates a possible difference between the intensities or phases of
both lasers, and can be used to create a dark--state. If the laser detuning is zero and if we choose
a value $A=-\langle m_{x}^{0}|e^{ikx}|m_{x}^{0}\rangle/\langle m_{y}^{0}|e^{iky}|m_{y}^{0}\rangle$, 
for a particular two--dimensional state $|m_{x}^{0},m_{y}^{0}\rangle$, then the selected level 
remains dark respect to the laser pulse \cite{1Atom}.
\end{itemize}
In absence of collisions, the above mentioned dark--state mechanisms 
\cite{1Atom,Manyatoms} 
allow to cool the atoms not only into the ground state, but
also into an arbitrary excited state of the trap. 
In order to achieve that one has to use different dynamical 
cooling schemes. Each dynamical cooling cycle 
must contain sequences of pulses of
appropriate frequencies.  The following types of pulses are employed: i) {\it confinement
pulses}: spontaneous emission may increase each of the quantum numbers
$m_{x,y,z}$ by $O(\eta^2)$. In $D$-dimensions pulses with detuning
$\delta = -D\hat \eta^{2} \omega$, where $\hat\eta^2$ is the closest
integer to $\eta^2$, have thus an overall cooling effect, and confine
the atoms in the energy band of $D$ recoils; 
ii) {\it dark-state cooling pulses}: these pulses should fulfill dark
state condition for a selected state to which the cooling should occur;
iii) {\it sideband and auxiliary cooling pulses}: in general, dark state
cooling pulses might lead to unexpected trapping in other levels. In
order to avoid it, auxiliary pulses that empty undesired dark states and
do not empty the desired dark state are needed; 
iv) {\it pseudo-confining pulses}: with the  use
of pulses i)--iii) cooling is typically very slow. In order
to shorten cooling time we use pulses with  $\delta = -3
\eta^{2}\omega/2$ and $\delta = -\eta ^{2}\omega$, which 
pseudo-confine the atoms below
$n=3\eta ^{2}/2$ and $n= \eta ^{2}$. 

\subsection{Numerical results}

\begin{figure}[ht]
\begin{center}\
\epsfxsize=7.0cm
\hspace{0mm}
\psfig{file=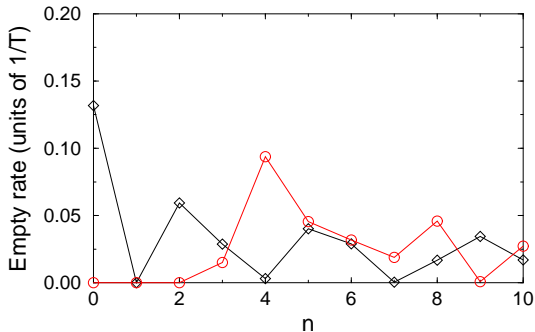,width=7.0cm}\\[0.1cm]
\caption{One--atom empty rates as a function of the trap level for the case of $eta=3.0$, and a laser pulse with 
detuning $\delta=s\omega$, with $s=8$ (diamonds) and $s=-3$ (circles).}
\label{fig:1}
\end{center}
\end{figure}

In the following we simulate the dynamics of the atomic population in the different trap levels 
using standard Monte Carlo methods. We consider the case of an harmonic trap 
with Lamb--Dicke parameter $\eta=3$. We assume $\gamma=0.04\omega$, 
$\Omega=0.03\omega$ (consequent with the Festina Lente limit), 
and a number of atoms $N=133$ (well in the 
weak--condensation regime). 
The calculations have been performed 
taking into account $40$ trap levels. As an initial condition, we assume in all the following graphics 
a thermal distribution with mean $\langle n \rangle =6$. In order to compare with the calculations without 
collisions, we analyse the same cooling scheme into the level $n=1$ of the trap, as that studied 
in Refs.\ \cite{Manyatoms}, for the case 
without collisions. We consider cycles of four laser pulses with detunings $\delta=s\omega$, where 
$s_{1,2,3,4}=-9,8,-10,-3$, and time duration $T=2\gamma/\Omega^2$. Pulses $1$ and $3$ are confining pulses, 
pulse $2$ is a dark--state pulse for $n=1$, and pulse $4$ is an auxiliary pulse. The one--atom emptying rates 
\cite{1Atom}, $|\langle n+s|\exp(ikx)|n\rangle|^2$, 
for the first $10$ levels of the trap are presented in Fig.\ \ref{fig:1}. As one can observe, 
the effect of the pulses 
is to empty all the states except $n=1$, which acts consequently as a trapping state. Observe, that due to the 
characteristics of the Franck--Condon factors, some levels of the trap are barely emptied, in particular for 
this case, $n=7$ is also a quasi--dark state for pulse $2$, and is poorly emptied by the auxiliary pulse $4$. 
This is not important in the case without collisions, because $n=1$ remains the darkest level throughout all 
the dynamics. We shall show in the following that this is no more true when the collions are accounted for.

% Figure 1

% Figure 2
\begin{figure}[ht]
\begin{center}\
\epsfxsize=5.0cm
\psfig{file=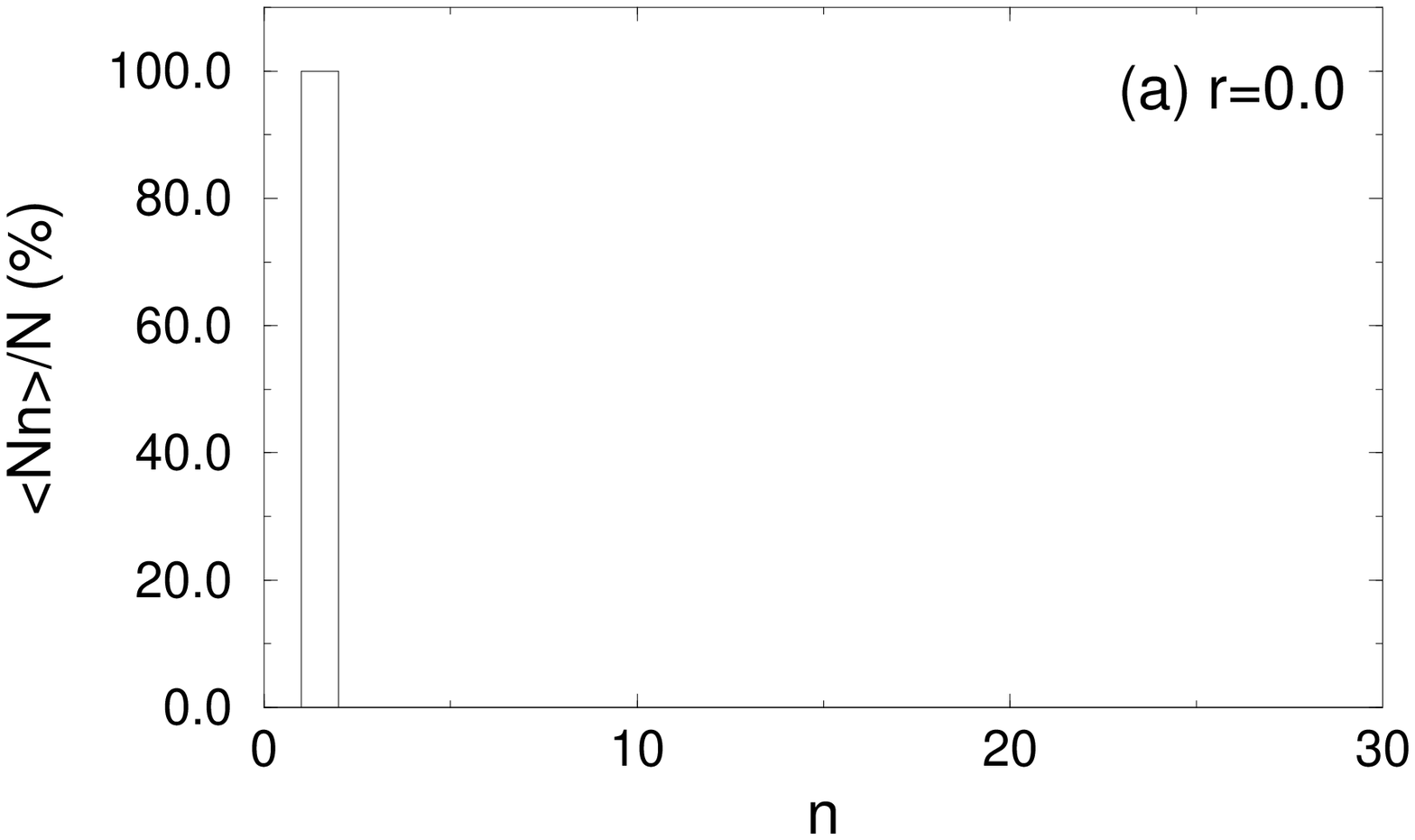,width=6.0cm}\\
\psfig{file=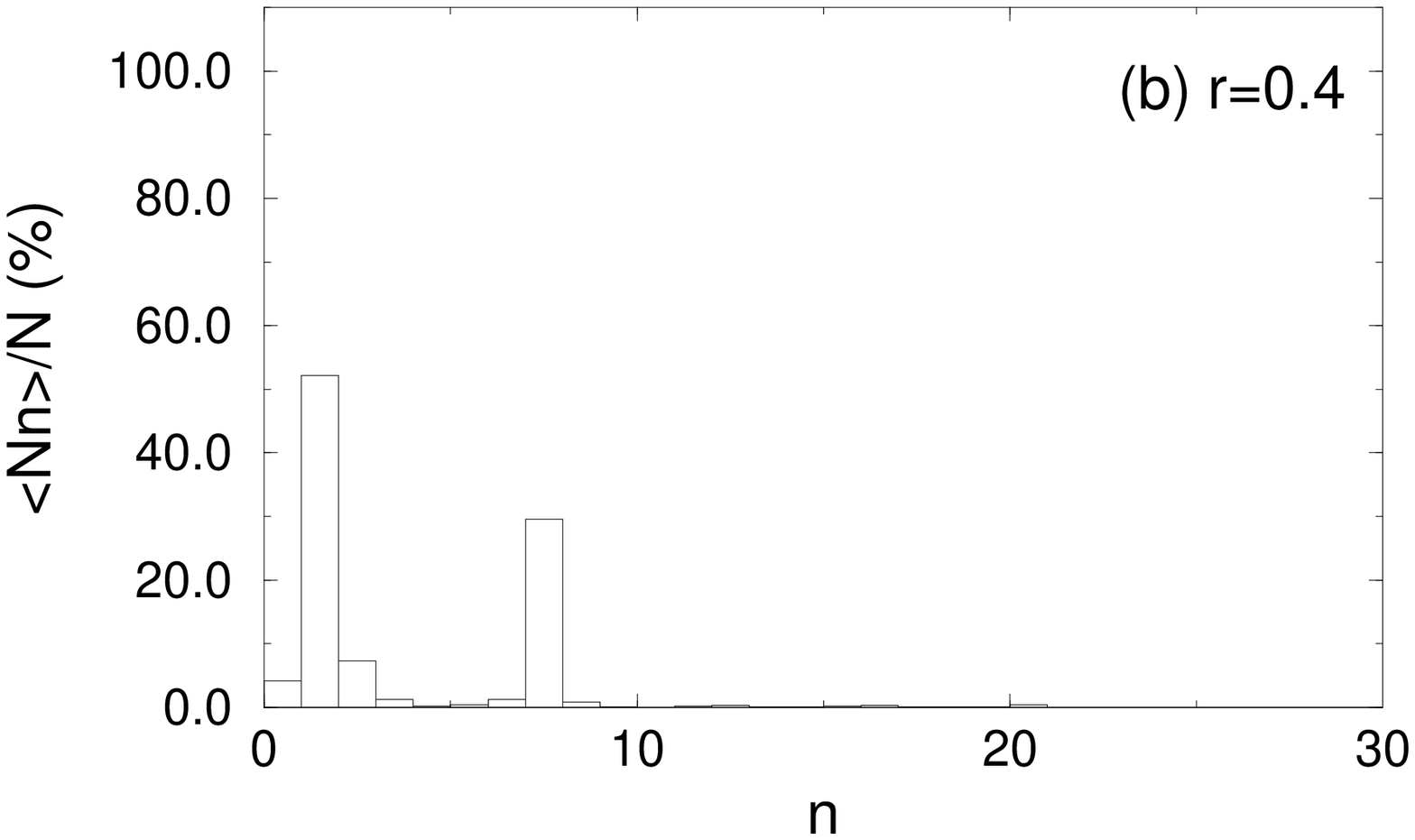,width=6.0cm}\\
\psfig{file=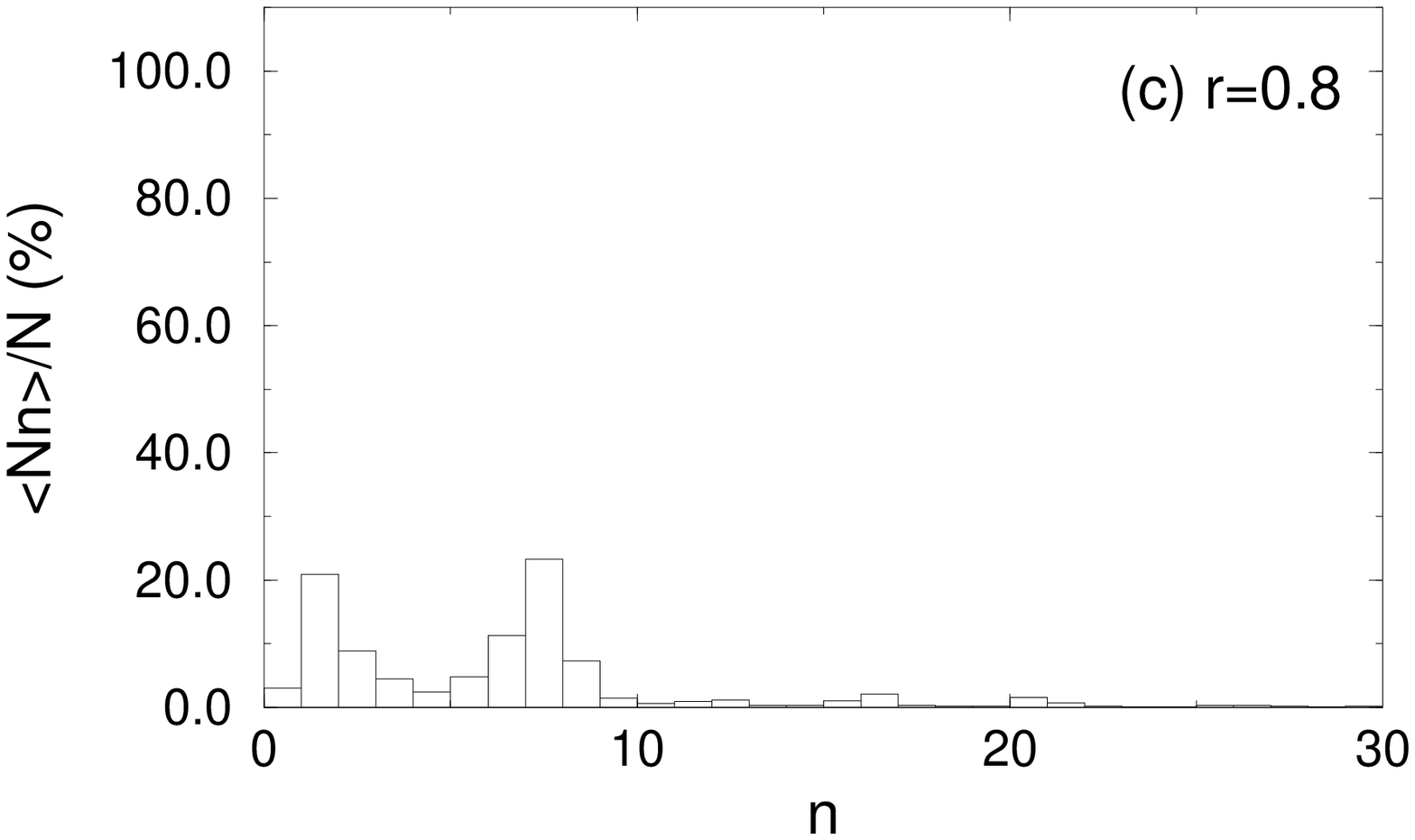,width=6.0cm}\\
\psfig{file=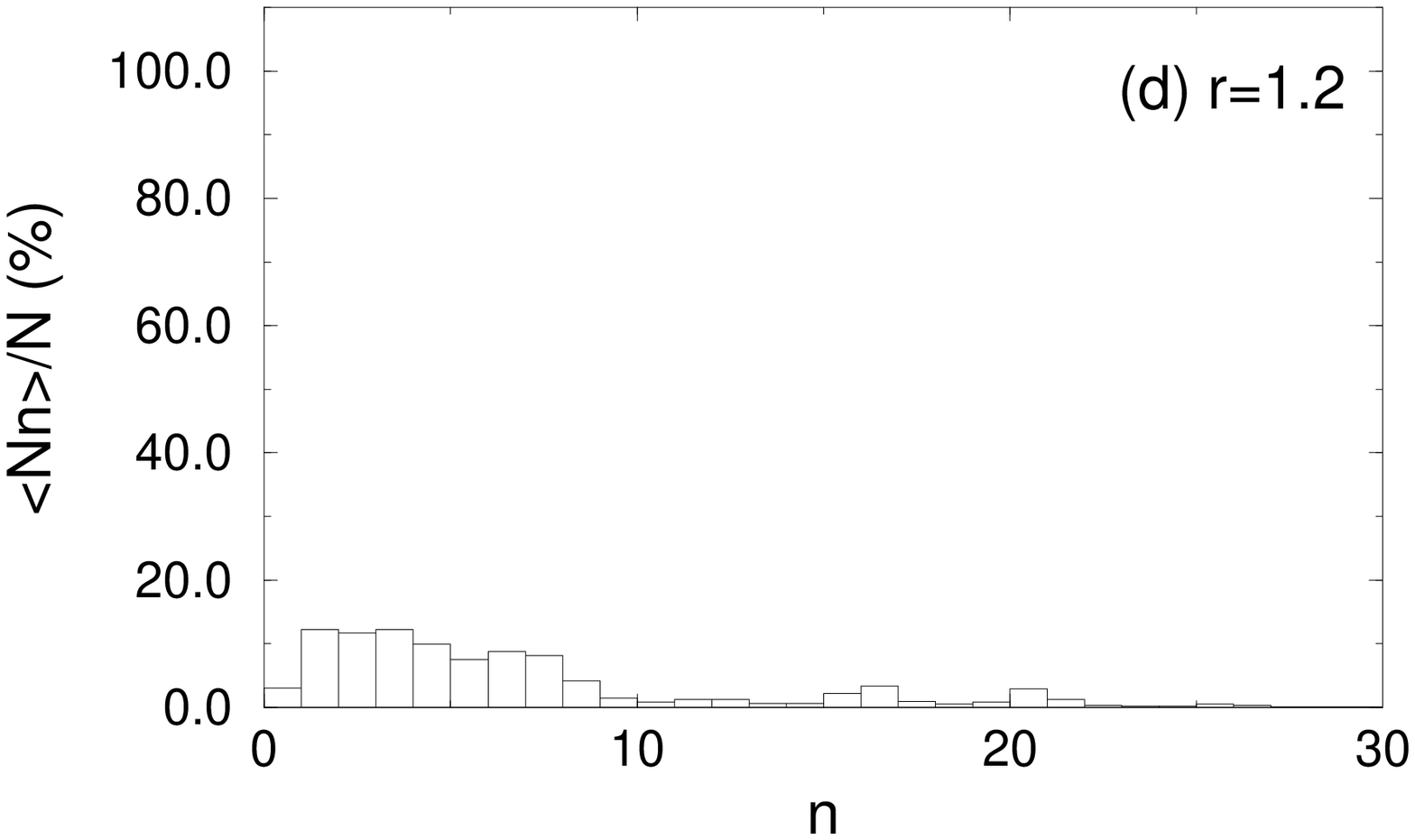,width=6.0cm}\\
\psfig{file=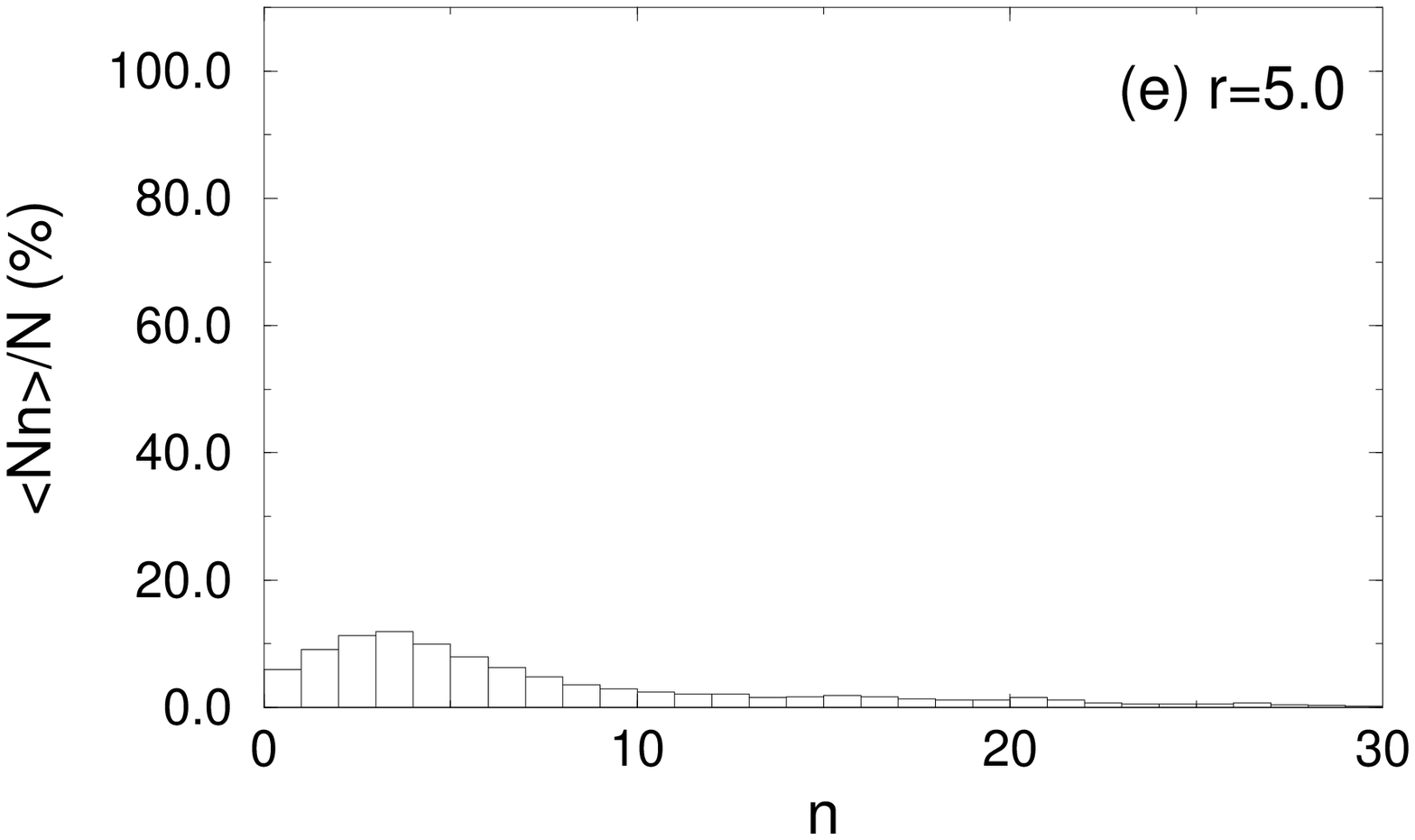,width=6.0cm}\\[0.1cm]
\caption{Averaged population of the trap levels, for the case of $\eta=3.0$ and cooling cycles consisting of 
four pulses with durations $T=2\gamma/\Omega^2$, and 
detunings $\delta=s\omega$ with $s=-9,8,-10,-3$. The cases of (a) r=0.0, (b) r=0.4, (c) r=0.8, (d) r=1.2 and 
(e) r=5.0 are considered.}
\label{fig:2}
\end{center}
\end{figure}

\begin{figure}[ht]
\begin{center}\
\epsfxsize=6.0cm
\hspace{0mm}
\psfig{file=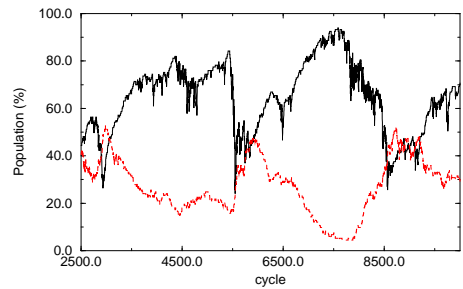,width=6.0cm}\\[0.1cm]
\caption{Detail of the dynamics of the populations of levels $n=1$ (solid line) and $n=7$ (dashed line), for the 
case of $\eta=3.0$, $r=0.4$. Each cooling cycle consists of four pulses with durations $T=2\gamma/\Omega^2$, and 
detunings $\delta=s\omega$ with $s=-9,8,-10,-3$. The non--linear pseudo--oscillation between $n=1$ and $n=7$ can 
be clearly observed.}
\label{fig:3}
\end{center}
\end{figure}

Fig.\ \ref{fig:2} shows the evolution of the averaged population of the level $n=1$ when $r$ grows. 
We have evolved the system under 5000 
cooling cycles (to avoid the effects of the initial conditions), and performed the average from the cycle $5000$ 
until the cycle $15000$. In Fig.\ \ref{fig:2} we have depicted the averaged population distribution for the cases of 
(a) $r=0$ (ideal gas), (b) $r=0.4$, (c) $r=0.8$, (d) $r=1.2$, and (e) $r=5.0$. For the ideal--gas case 
$r=0$ one obtains that 
the population is completely condensed into the level $n=1$ \cite{Manyatoms}. When $r$ is increased the 
laser--induced condensation into the excited state $n=1$ is destroyed, but in a non--trivial way. We observe in 
Fig.\ \ref{fig:2} (b) that for $r=0.4$ the population is basically distributed 
in two well defined peaks, one in $n=1$ and 
the other in $n=7$. The reason for this behavior can be understood very well using Fig.\ \ref{fig:1}. In absence of 
collisions the level $n=1$ is not emptied at all, while, as pointed out previously, $n=7$ is emptied, but slowly, 
and therefore at the end the population is finally transferred to the level $n=1$. However in the presence of 
collisions, level $n=1$ is still a dark--state for the laser, but it is emptied by the collisions 
with a frequency proportional to $N_1^2$. This means that the population is pumped into $n=1$ due 
to the laser cooling, but the more the population we pump into $n=1$ the more the level is emptied via collisions.
This effect can be well illustrated by Fig.\ \ref{fig:3}, where one can observe 
periods of filling of $n=1$ followed by 
abrupt decays of the level population. The emptying of level $n=1$ is mainly produced via collisions between 
two atoms in the level $n=1$ to produce two atoms in $n=0$ and $n=2$ respectively. The laser cooling provides a 
mechanism to repump such expelled population from the level $n=0$ and $n=2$ back to the level $n=1$. Such control is 
already maintained for large occupations of $n=1$, but in an unstable way, due to the highly non--linear character 
of the dynamics. A slight excess of population into the level $n=0$ and $n=2$ provoques a speed--up of the emptying 
process of $n=1$. This situation is reflected, for example, in the behavior of the system between cycles 
$4500$ and $5500$ in Fig.\ \ref{fig:3}. In particular, level $n=1$ can become more emptied than $n=7$, 
and the latter turns to be the effective darkest 
level, i.e. the level less emptied. Therefore the population tends to be transferred into $n=7$. 
But, when $N_7$ increases 
so does the empty rate of the level $n=7$, which can become larger than that of $n=1$, and so on. Therefore, as 
consequence of this process a non--linear pseudo--oscillatory motion between the populations 
of $n=1$ and $n=7$ is produced, 
as observed in Fig.\ \ref{fig:3}. 
This oscillatory motion leads to the two--peaked distribution of Figs.\ \ref{fig:2}. Finally, when 
$r$ becomes very large the collision dynamics is much faster than the cooling time, and the peaked structure 
dissapears, as observed in Figs.\ \ref{fig:2} (c), (d) and (e). 
Observe that nevertheless the effects of the laser cooling 
mechanism are nevertheless present in Fig.\ \ref{fig:2} (f). In absence of laser cooling, 
it can be demonstrated that for 
$r=5$, the population has a maximum in $n=0$. On the contrary, in the presence of the laser, 
the population of $n=0$ is very efficiently 
and rapidly emptied by the pulse with detuning $s=8$, which is the most rapid cooling process. For larger $r$
even this process is eventually overcome, 
and one recovers the same distribution as that obtained only considering the 
collisions without laser cooling.

\section{Three-dimensional results}
\label{sec:3D}

In this section we analyze the case of the laser--cooling into the ground state of
an isotropic three--dimensional harmonic trap, of frequency $\omega$. The numerical calculation of the system 
dynamics for the three--dimensional case is quite complicated, due to both the degeneracy of the levels, and the 
difficulties to obtain reliable values for the integrals $U(n_1,n_2,n_3,n_4)$. Therefore, we shall limit 
ourselves to the use of the ergodic approximation, i.e. we shall assume that states with the same energy are 
equally populated. The populations of the degenerate energy levels equalize on a time scale much faster than the 
collisions between levels of different energies, and than the laser--cooling typical time. This approximation 
leads to the correct steady--state distribution, although the dynamics can 
be slightly different than in the non--ergodic calculation \cite{QK2}. 
% Figure 3

Following ref.\ \cite{Holland} the probability of a collision of two atoms in energy shells $n_1$ and $n_2$, to 
give two atoms in shells $n_3$ and $n_4$ (where this collision is assumed to change the energy distribution 
function), is of the form:
\begin{eqnarray}
&&P(n_1,n_2\rightarrow n_3,n_4)=\Delta (n_j+1)(n_j+2) \nonumber \\
&&\times\frac{N_{n1}(N_{n2}-\delta_{n_2,n_1})(N_{n_3}+g_{n_3})(N_{n_4}+g_{n_4}+\delta_{n_3,n_4})}
{g_{n_1}g_{n_2}g_{n_3}g_{n_4}},
\end{eqnarray}
where $g_{n_k}=(n_k+1)(n_k+2)/2$ is the degeneracy of the energy shell $n_k$, 
$n_j=min \{ n_1,n_2,n_3,n_4 \}$, and $\Delta=(4a^2\omega^2m)/(\pi\hbar)\sim 1.5\times 10^{-5}$. 
Concerning the laser--cooling probabilities we shall use the same expressions as 
those already developed in Refs.\ \cite{Manyatoms}.

\begin{figure}[ht]
\begin{center}\
\epsfxsize=6.0cm
\hspace{0mm}
\psfig{file=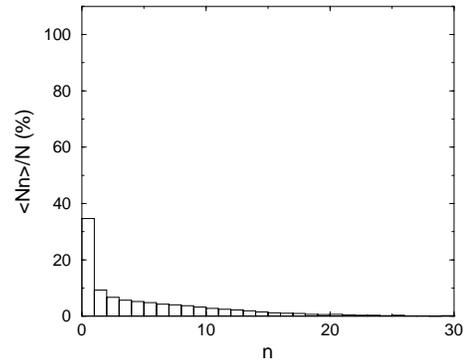,width=6.0cm}\\[0.1cm]
\caption{Average of the population of the different energy shells, after evolving the system 
under collisions, starting from a themal distribution of $\langle n \rangle=6$.}
\label{fig:4}
\end{center}
\end{figure}

\begin{figure}[ht]
\begin{center}\
\epsfxsize=6.0cm
\hspace{0mm}
\psfig{file=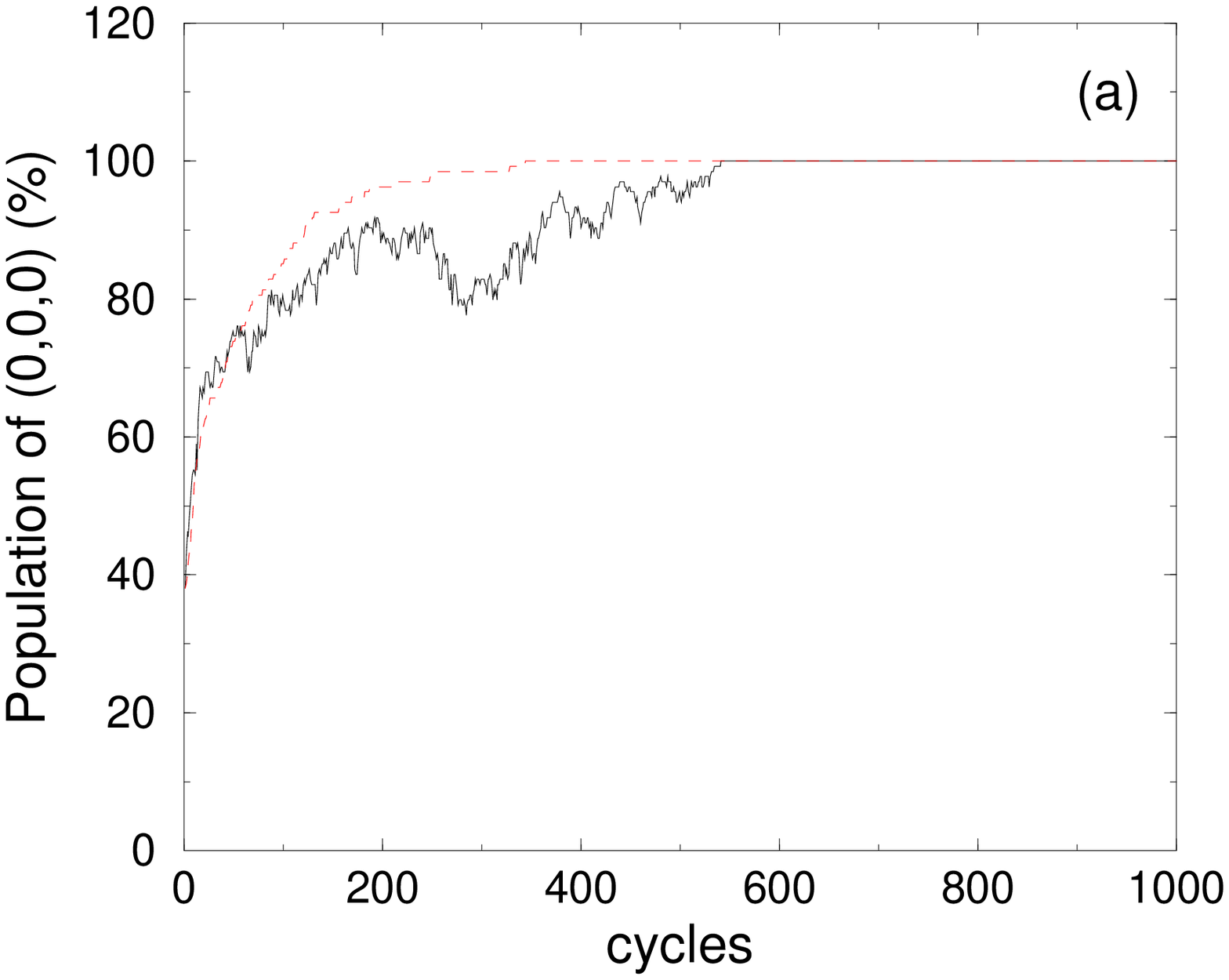,width=6.0cm}\\[0.1cm]
\psfig{file=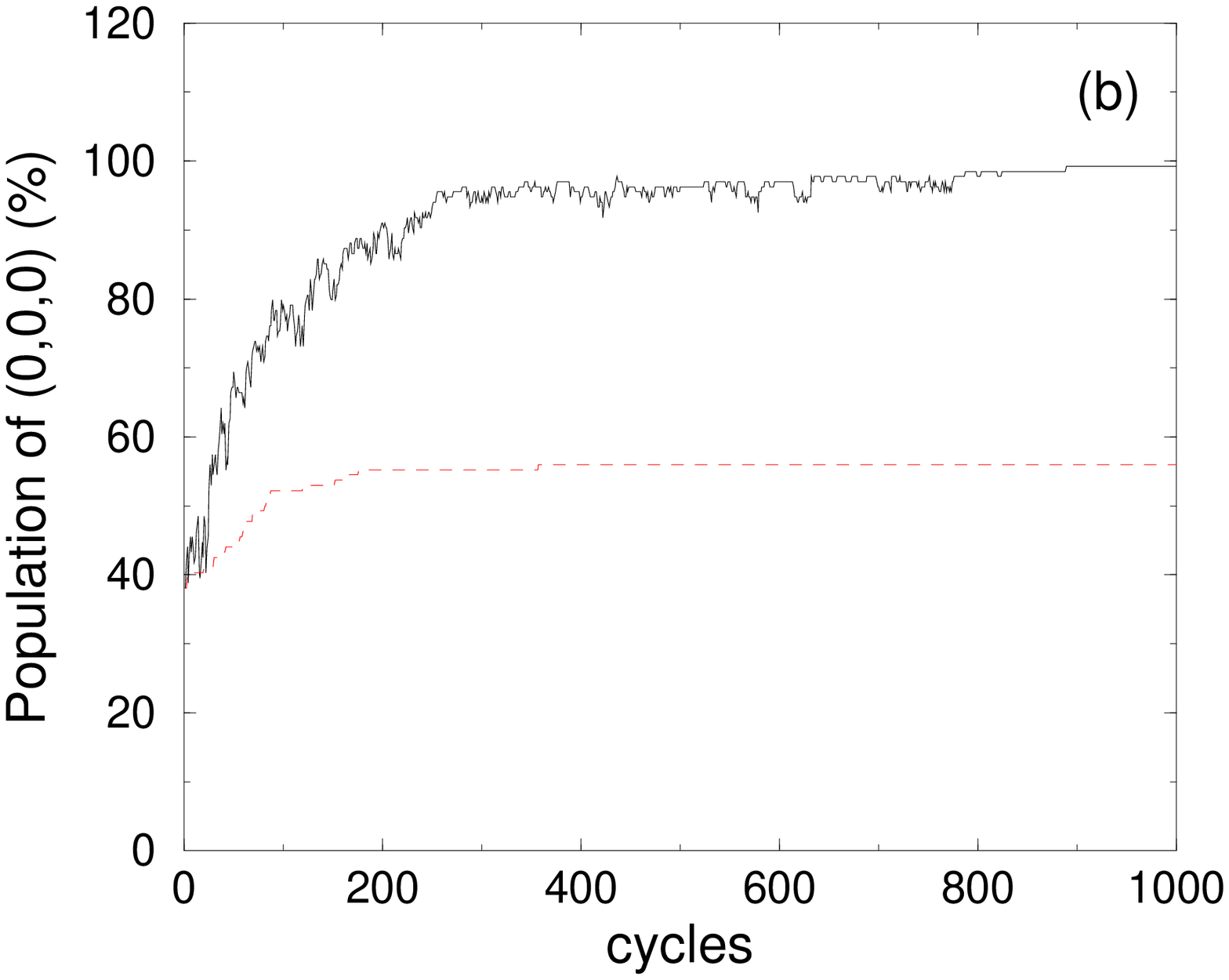,width=6.0cm}\\[0.1cm]
\caption{Dynamics of the population of the ground state of the trap for the case of
$\eta=2$, $\gamma=0.04\omega$, $\Omega=0.03\omega$, $N=133$, starting from the BED distribution of 
Fig.\ \ref{fig:4}. (a) Cooling cycles composed of two laser pulses of detunings 
$\delta=s\omega$, with $s_{1,2}=-4,0$, and $A_z=1,-2$ 
repectively, are used. The simulations taking into account the collisions (solid line) and no taking them 
into account (dashed line) are compared. (b) Same case, but only employing pulse $1$.}
\label{fig:5}
\end{center}
\end{figure}

In the following we simulate the evolution of the system by using again Monte Carlo simulations. Due to numerical 
limitations we consider a Lamb--Dicke parameter $\eta=2$. We assume as previously $\gamma=0.04\omega$ and 
$\Omega=0.03\omega$, and a number of atoms $N=133$. As a first step, we begin with a thermal distribution of 
mean $\langle n\rangle=6$, and evolve the system just with collisions, until obtaining a Bose--Eintein distribution 
(BED) (which does not coincide exactly with the 
thermodynamical one, due to finite--size effects), see 
Fig.\ \ref{fig:4}. The distribution obtained in this initial step serves as the initial state for laser 
cooling. As we see, it already contains quite subtantial amount of atoms condensed in the ground state, but also 
a lot of uncondensed ones. Laser cooling will transfer the latter ones into the ground state.
We apply our laser cooling cycles, each one of them composed by two laser 
pulses of detuning $\delta=s\omega$, with $s_{1,2}=-4,0$, and time duration 
$T=(2\gamma)/\Omega^2$. The laser pulses are emitted in three orthogonal directions $x$, $y$ and $z$, and are 
characterized by their respective rabi frequencies $\Omega_x=\Omega_y=\Omega$, $\Omega_z=A_z\Omega$. For the first 
pulse we assume $A_z=1$, while for the second one $A_z=-2$ is considered. With this choice, the second pulse is 
an "interference"--dark--state pulse for the ground--state of the trap. Fig.\ \ref{fig:5}(a) shows 
(dashed line) that these two 
pulses are able to condense the population into the ground state of the trap, in absence of collisions; 
in particular no confinement pulses (of detunings $\delta=-12\omega$ in this case) are needed. This is due to the 
bosonic enhancement and the fact that initially the system is already partially condensed. The dark--state pulse 
is neccesary to repump the population in those states of the energy shells $1$, $2$ and $3$, 
which are dark respect to the pulses with detuning $\delta=-4\omega$. Fig.\ \ref{fig:5} shows 
(solid line) the dynamics of the 
population of the ground-state in presence of collisions. After 600 cycles, 
all the population is transferred to the ground state of the 
trap. This means that applying the laser cooling scheme brigs the system into an 
effective BED of $T=0$. It is easy to undertand why the effect is maintained in presence of collisions, 
even considering that the collisional dynamics is much faster than the laser--cooling one.
The laser--cooling mechanism tends to 
decrease the energy per particle (i.e. the chemical potential of the system), in the same way as evaporative 
cooling does, but without the losses of particles in the trap during the process. Thermalization via 
collisions brings the system to a lower temperature. Repeating the laser cooling sufficient times the system 
ends with an effective zero temperature. Finally, let us point out that some auxiliary pulses which are needed 
in the ideal gas, are not in presence of collisions. In particular for the previous example, the pulse of zero 
detuning (required for the ideal gas case, Fig.\ \ref{fig:5}(b) dashed line) is no more needed, 
as shown in Fig.\ \ref{fig:5}(b) (solid line). Thus, the 
laser--cooling scheme is not only possible in presence of collisions, but can be even significanly simplified.

% Figure 4
% Figure 5

\section{Conclusions}
\label{sec:conclu}

In this paper, we have analysed the effects of the atom--atom collisions on the colective laser cooling
of bosonic gases trapped in an harmonic trap, under the Festina--Lente condition. 
In particular, we have studied the case in which the mean--field energy provided 
by the atom--atom collisions is much smaller than the typical energy of the harmonic trap.
Under such conditions, we have derived the ME which describes the system, and observed that such ME 
splits into two parts:(i) a purely collisional part which has the form of a QBME, and (ii) a purely 
laser--cooling part, which has the same form as the ME which describes the laser--cooling in absence of 
collisions. By using this ME, we have simulated the dynamics of the trapped gas for different situations.
First, we have analysed the cooling into an excited state of a one--dimensional trap. 
We have observed that the transition from the ideal--gas limit (in which the atoms are completely condensed 
into the chosen excited state) to the case in which the collisions dominate the dynamics, is not trivial, 
specially when the collisional and cooling time scale are comparable. 
In such a case, cooling and collisional 
processes enter in competition, and new phenomena can appear, as 
for example unstable population of 
an excited state followed by abrupt population decays, and non--linear pseudo--oscillations 
between different trap levels. We have finally analyzed the laser--cooling into the ground state of an 
isotropic three--dimensional harmonic trap, by using the ergodic 
approximation. We have shown that, although the collisional time is typically much faster 
than the cooling time scale, the laser cooling allows to transform a BED with a finite temperature into 
an effective BED with zero temperature. The laser cooling reduces 
the chemical potential of the trapped atoms, while 
the collisions provide the thermalization.

Let us finally present important remarks concerning the scaling of our theory, the situation beyond the 
weak--condensation regime, and the problem of the two- and three--body losses in the trap.
First, we
stress  that we have presented here the results  obtained for $\eta=2,3$ only for 
the reasons of numerical complexity which grows rapidly with $\eta$. 
Qualitatively, the same results can be obtained for larger  $\eta$'s, and
therefore, for lower densities. In fact, we have observed similar results for
$\eta=5$ in  one dimensional simulations. If we increase $\eta$ by factor $F$, the
corresponding density (for fixed $N$) decreases as  $F^{-3}$, the three
body loss rates as $F^{-6}$, whereas the trap frequency decreases as
$F^{-2}$, which means that the corresponding cooling time (to fulfill the
{\it Festina Lente} conditions) will increase as $F^2$, i.e. much less
than the lifetime due to three-body collisions. 
 
Beyond the weak condensation regime, the mean--field energy cannot be neglected, 
and therefore the trap levels are no more the harmonic ones. This 
has a two--fold 
consequence: (i) The levels of the trap are non--harmonic, i.e. they are not equally 
separated, because their energies become dependent on the occupation numbers; (ii) 
the wavefunctions are different, and in particular the condensate wavefunction 
becomes broader (we consider here only the case of repulsive interactions, 
$a>0$). The fact that the energy levels 
are not harmonic any more, complicates the laser cooling, but the use of pulses 
with a variable frequency and band--width should produce the same  results 
as those presented here. 
The point (ii) implies that the central density of the interacting gas 
is much lower than the one predicted for noninteracting particles. In fact, the ratio between 
the interacting--gas central density (in Thomas--Fermi (TF) approximation) 
and the ideal--gas central density, goes as \cite{Stringari}
\begin{equation}
\frac{n_{TF}}{n_{ideal}}=\frac{15^{2/5}\pi^{1/2}}{8}\left ( \frac{Na}{a_{HO}} \right )^{-3/5},
\end{equation}
where the central density for the ideal case is given by $n_{ideal}=N/(\pi^{3/2}a_{HO}^3)$. 

The above result has important consequences, when 
one considers the problem of three--body collisions, which usually 
begin to play a role at densities of the order of $10^{15}$ atoms/cm$^3$.
For example, let us analyse the case of Sodium, for which 
$a_R=\sqrt{\hbar/m\omega_R}=0.132 \mu {\rm m}$, and $a=2.75${\rm nm}. From the definition 
of $\eta$, $a_{HO}=\eta a_R$.  
For the ideal gas case, the regime in which three--body losses are important 
is reached for $N\simeq 12.8 \eta^3$. For $\eta=8$ this means $N=6.5\times 10^3$. Amazingly, 
for the interacting gas, the same is true for $N \simeq 5.1 \eta^6$, and therefore for $\eta=8$, 
the regime in which three--body losses are important 
is reached for $N=1.3\times 10^6$. Below this number, 
the interaction between the particles is dominated by the ellastic two--body collisions 
considered in this paper. As point out above,  our laser cooling scheme could be extended 
beyond the weak--condensation regime, and therefore laser--induced condensations of more than 
$10^6$ atoms are feasible. 

Concerning other loss mechanisms, we have to mention here the hyperfine changing two-body collisions,
or generally speaking any inelastic two-body processes. These can be supressed completely if we cool
atoms into the absolute ground internal state, which is possible in the dipole traps. For alkalis
this is typically  done by cooling to the lowest energy state in the 
lower hyperfine manifold  (external static magnetic
fields are used  to split the levels within the hyperfine manifold). 

Finally, yet another loss mechanism disregarded here is due to photoassociation, i.e. excitation of
molecular resonances. This kind of loss rates  are typically
of the order $\gamma n(\lambda/2\pi)^3$ (where $\gamma$ is the linewidth of the auxiliary level $|r\rangle$, 
$\lambda$ is the laser wavelength, and $n$ is the atomic density), i.e. 
allow for achieving about $1000$ cooling cycles of duration $\simeq 1/\gamma$ provided the density
remains smaller than $10^{12}-10^{13}$ atoms/cm$^2$. However, the photoassociation losses can be 
reduced by several  orders of magnitude if the laser is red detuned,
 and tuned exactly in the 
middle of the molecular resonances \cite{shlypm} (note that this is the detuning respect to the one--photon 
transitions from the ground states to the state $|r\rangle$, and not the two--photon detuning). 
Other, possibility, is of course to use 
a more intense laser tuned below the Condon point, i.e. the minimum of the molecular potential.

We acknowledge  support from Deutsche Forschungsgemeinschaft (SFB
407) and the EU through the TMR network ERBXTCT96-0002. 
We thank J. I. Cirac, Y. Castin, G. Birkl, K. Sengstock, W. Ertmer, T. Pfau and T. Esslinger for 
fruithful discussions.

\end{document}